\documentclass{article}
\usepackage{latexsym}
\usepackage{amsfonts}
\usepackage{amssymb}

\def\abstract#1{\vskip 7mm 
        \begin{center}{\large Abstract}\par \smallskip
                \begin{minipage}[c]{12cm}
                        \small #1
                \end{minipage}
        \end{center}
}
\def\title#1{\begin{center}{\Large\bf #1}\end{center}}
\def\author#1{\vskip 5mm \begin{center}{#1}\end{center}}
\def\address#1{\begin{center}{\it #1}\end{center}}
\makeatletter

\newcommand{\order}{{\cal O}}
\newcommand{\beq}{\begin{equation}}
\newcommand{\eeq}{\end{equation}}
\newcommand{\beqa}{\begin{eqnarray}}
\newcommand{\eeqa}{\end{eqnarray}}
\newcommand{\bea}{\begin{eqnarray}}
\newcommand{\eea}{\end{eqnarray}}
\newcommand{\mpl}{M_{Pl}}
\newcommand{\mg}{M_{G}}

\newcommand{\rhov}{\rho_{\rm v}}

\newcommand{\rhocrz}{\rho_{\rm cr0}}

\newcommand{\omegamz}{\Omega_{\rm m0}}
\newcommand{\omegabz}{\Omega_{\rm b0}}
\newcommand{\omegavz}{\Omega_{\rm v0}}

\@ifundefined{lesssim}{\def\lesssim{\mathrel{\mathpalette\vereq<}}}{}
\@ifundefined{gtrsim}{\def\gtrsim{\mathrel{\mathpalette\vereq>}}}{}
\def\vereq#1#2{\lower3pt\vbox{\baselineskip1.5pt \lineskip1.5pt
\ialign{$\m@th#1\hfill##\hfil$\crcr#2\crcr\sim\crcr}}}
\makeatother

\begin{document}

\title{%
  Issues on the cosmological constant
}
\author{%
 Jun'ichi Yokoyama
}
\address{%
  Department of Earth and Space Science,
  Graduate School of Science, \\
 Osaka University, Toyonaka 560-0043
}

\abstract{
Some issues of the cosmological constant or dark energy are briefly reviewed.
There are an increasing number of observations that constrain 
the equation of state of dark energy more stringently
and favor the time-independent cosmological constant.
Then a plausible model of dark energy would be a theory with
degenerate perturbative vacua in which its origin is explained by a
nonperturbative effect so that, unlike quintessence, k-essence etc.,
 it is separable from the perturbative problem
why its amplitude is smaller than the Planckian density by a factor
of $\order (10^{-120})$.
}

\section{Introduction}

Originally the cosmological constant $\Lambda$ was introduced as
an undetermined
constant in the left-hand-side of the Einstein equation,
\beq
R_{\mu \nu }  - {\textstyle{1 \over 2}}Rg_{\mu \nu }  + \Lambda g_{\mu
\nu }  = 8\pi GT_{\mu \nu }, 
\eeq
so that this equation had a static cosmological solution,
guided by the strong prejudice that our Universe is static \cite{Ein}.
As is well known, however, the solution thus obtained is very unstable
and is not realistic.  As a natural consequence, it turned out that we
live in a dynamically expanding universe as theoretically pointed out by
Friedmann and observationally discovered by Hubble \cite{Hubble}.

Nowadays it is more appropriate to put $\Lambda$ in the 
right-hand-side of the Einstein equation as a part of the energy-momentum
tensor.
By nature the cosmological constant is equivalent to the vacuum energy
density $\rhov$ with the following relation.
\beq
\rho _{\rm{v}}  = \frac{\Lambda }{{8\pi G}} 
= \frac{{M_{Pl}^2 }}{{8\pi }}\Lambda  = M_G^2 \Lambda, 
\eeq
where $\mpl$ and $\mg\equiv\mpl/\sqrt{8\pi}=2.4\times 10^{18}$GeV are
the Planck mass and the reduced Planck scale, respectively. 

In the field theoretical point of view, if we 
add the zero-point fluctuation of a quantum field up to a cutoff $k_c$,
we find
\beq
\left\langle {\rho _v} \right\rangle =\int_0^{k_c} {{{d^3k} \over
{\left( {2\pi } \right)^3}}}{1 \over 2}\sqrt {k^2+m^2}\cong {{k_c^4}
\over {16\pi ^2}}. \label{qft}
\eeq
Taking the cutoff $k_c$ at the Planck scale beyond which classical
gravity may not be used, we find $\langle\rhov\rangle=4\mg^4$.  Thus the
natural scale of the vacuum energy is the Planck scale \cite{Weinberg}.

Observationally cosmic energy density today is at most of order of the
critical density $\rhocrz\cong 10^{-120}\mg^4$.  
Confronting such a huge discrepancy of 120 digits between the theoretically
natural value and the observationally allowed upper bound, it was
considered for a long time that there exist some mechanism to make
$\Lambda$ exactly vanish and its quest  is the conventional
cosmological constant problem to which we refer the Problem I.

Recent advance of observational cosmology, however, has revealed that
our Universe is in a stage of accelerated expansion now 
\cite{SN}.  This
means that there exists a tiny magnitude of cosmological
constant and/or some unknown form of matter with negative pressure
driving accelerated expansion which is generically termed as
 dark energy.  To account for the origin of this small dark energy is
 the new cosmological constant problem referred to the Problem II.

In this talk I first summarize the currently favored observational
values of the cosmological parameters and then move on to Problems I and
II and approaches to their solutions.  Finally I mention the expected
future of our Universe.

\section{Cosmological parameters}

Current trend of observational cosmology is that the values of
cosmological parameters are converging to that of the so-called
concordance model.
Although  the forthcoming data of MAP satellite will certainly provide
more precise informations on them, it is still remarkable that
combinations of many
different means of observational determination of cosmological
parameters have already terminated most of the
long-standing disputes on them.

\subsection{Hubble parameter}
The key project of the Hubble Space Telescope has determined the value
of the Hubble parameter out to about 400Mpc with various secondary
indicators based on the primary cephid distance which gives $H_0=75\pm
10$km/s/Mpc.  The results of the secondary indicators are summarized as
follows. 
\begin{center}
\begin{tabular}{ll}
Type Ia Supernovae & $H_0=71\pm 2({\rm stat})\pm 6({\rm syst})$km/s/Mpc
 \\
Tully-Fisher relation &  $H_0=71\pm 3({\rm stat})\pm 7({\rm syst})$km/s/Mpc
 \\
Surface brightness fluctuations &  $H_0=70\pm 5({\rm stat})\pm 6({\rm syst})$km/s/Mpc \\
Type II Supernovae &  $H_0=72\pm 9({\rm stat})\pm 7({\rm syst})$km/s/Mpc
\\
Fundamental plane of elliptical galaxies &  $H_0=82\pm 6({\rm stat})\pm 9({\rm syst})$km/s/Mpc
\end{tabular}
\end{center}
Combining these results Freedman et al.\ conclude that the final result
of HST key project is $H_0=72\pm 8$km/s/Mpc \cite{hst}.

Sunyaev-Zel'dovich effect provides other means of determination of
$H_0$, although it suffers from fairly large systematic errors.
The current result is reported as $H_0=60\pm4^{+13}_{-18}$km/s/Mpc, which is
consistent with the HST result \cite{SZ}.

\subsection{Matter density}
Nonrelativistic matter consists of baryon and cold dark matter whose
identity is still unknown but two primary candidates are neutralinos and
axions.  Observation of light elements and standard big-bang
nucleosynthesis (SBBN) gives $\Omega_{\rm b0}=(0.019-0.025)h^{-2}$ with
$h=H_0/100$km/s/Mpc \cite{Olive}.

There are a number of independent observational constraints on the
matter density $\omegamz$.  To name a few, the luminosity density and
average mass-to-light ratio of galaxies gives $\omegamz=0.19\pm0.06$
\cite{Carlberg},  while 
cluster baryon fraction from X-ray emissivity supplemented by 
$\omegabz$ from
SBBN yields $\omegamz=0.35\pm 0.07$ \cite{White}.  
On the other hand, shape parameter
of the transfer function of CDM scenario of structure formation,
$\Gamma=\omegamz h$ is fit by observation for $\Gamma\simeq 0.25$ which
means $\omegamz\simeq 0.35$ for $h\simeq 0.7$ \cite{Dodds}.  
Although we cannot pin
down the value of $\omegamz$ yet, currently favored value is around
$\omegamz\simeq 0.3$.

\subsection{Spatial curvature}
The angular power spectrum of cosmic microwave background radiation
(CMB) provides a unique probe for the spatial geometry of the Universe.
In particular, the location of the first Doppler peak reflects the
angular scale subtending the sound horizon at decoupling which is
sensitive to spatial curvature $K$.  Although the data before WMAP
suffer from fairly large errors around the first peak \cite{cc}, they favor
spatially flat universe \cite{Jaffe:2000tx}.
This is a good news because it is
 in accordance with standard inflation models. 
On the other hand, it raises a question: what makes up the rest of
the Universe, $1-\omegamz\simeq 0.7$?  The answer is, of course, dark
energy as we see below.

\subsection{Vacuum energy density}
Observations of 
the magnitude-redshift relation of high-redshift type Ia supernovae has
probed the deceleration parameter $q_0=(\omegamz-2\omegavz)/2$ in addition
to $H_0$ and they have found that $q_0$ is negative or our Universe is
in a stage of accelerated expansion \cite{SN}, 
which means that significant amount
of vacuum-like energy density exists today as
\beq
  \omegavz\simeq 1.25\omegamz+0.5\pm0.5.
\eeq

\subsection{The concordance model}
Combining all the above observational data 
we arrive at a model with $\omegamz\simeq 0.3$, 
$\omegavz\simeq 1-\omegamz\simeq0.7$.  Then the plausible range of the
cosmic age lies $t_0=(0.9-1.0)H_0^{-1}$.  The Hubble time corresponding
to $H_0=72\pm8$km/s/Mpc reads $H_0^{-1}=12.2-13.6-16.9$Gyr, which in
turn means $t_0=11-17$Gyr with the most likely value around $t_0\simeq
13$Gyr.  

Astrophysically, age of globular clusters gives $t_0=11-14$Gyr while
cosmological nuclear chronology yields $t_0=12-15$Gyr.  Ever since 
Hubble's determination of the Hubble parameter \cite{Hubble}, 
$H_0\sim 500$km/s/Mpc,
we confronted with the cosmic age problem from time to time in the
history of cosmology. The above set of cosmological parameters gives
consistent cosmic age with astrophysical observations and indeed 
deserves the
name of the concordance model.

Looking back more recent history of cosmology, only one and half decades
ago, theorists typically expected that we live in the Einstein-de Sitter
universe with $\omegamz=1$, $\omegavz=0$, $K=0$, and $H_0\sim
50$km/s/Mpc.  On the other hand, observational astronomers believed that
we live in an open universe with $\omegamz \lesssim 0.2$, $\omegavz=0$,
and $H_0=50$ or 100km/s/Mpc depending on schools \cite{S,dV}.  
Thus both parties can
compromise if they admit nonvanishing vacuum energy density taking an
intermediate value of the Hubble parameter.  A good news with the
concordance model is that it supports a theoretical prejudice that the
Universe is spatially flat and so is consistent with the prediction of 
standard inflationary cosmology, whose another prediction, generation of
almost scale-invariant adiabatic fluctuations, has also been supported
by observations of CMB anisotropy.  Thus we can describe evolution of the
Universe in the framework of inflationary cosmology.
Then the greatest mystery in cosmology is the questions, what is the
origin of the vacuum energy?  Why is it so small?

\subsection{Note added: WMAP results}
On February 12, 4AM (JST), the first-year result of the
Wilkinson Microwave Anisotropy Probe was disclosed \cite{wmap}.
As for the values of the cosmological parameters they basically 
 confirmed those of the concordance model with significantly smaller
 error bars than before  with $h=0.71^{+0.04}_{-0.03}$, 
$\Omega_{\rm b0}h^2=0.0224\pm 0.0009$, 
$\omegamz h^2=0.135^{+0.008}_{-0.009}$,
$\omegavz=\Omega_{\rm tot0}-\omegamz$, and  
$\Omega_{\rm tot0}=1.02\pm0.02$.  One should note, however, that the WMAP
data alone could be fit by other sets of parameters whose values are
quite different from those of the
concordance model.  The cosmic concordance is  achieved as a
result of combinations of various cosmological observations.

\section{Proposed solutions to the Problem I}
\subsection{Symmetry}
Whenever we encounter an unnaturally small quantity, we usually
interpret it as realized by virtue of some symmetry reason.
For example, in order to stabilize the Higgs' mass in the standard
particle physics model against quantum correction up to grand
unification or Planck scales, we usually introduce supersymmetry (SUSY).

As is well known, global SUSY also predicts vanishing vacuum energy
density at the supersymmetric ground state.  The SUSY generator,
$Q_\alpha$, satisfies the following anti-commutation relation.
\beq
\left\{ {Q_\alpha ,Q_\beta ^\dag} \right\}=\left( {\sigma _\mu }
\right)_{\alpha \beta }P^\mu ,
\eeq
with $P^\mu$ being the four momentum.  Since the supersymmetric ground
state is annihilated by this operator, we inevitably find vanishing 
energy density there.
\beq
Q_\alpha \left| 0 \right\rangle =Q_\alpha ^\dag\left| 0 \right\rangle
=0\Rightarrow \left\langle 0 \right|P^\mu \left| 0 \right\rangle =0. 
\label{susy}
\eeq
Unfortunately, this does not help to solve the Problem I at all,
because we know that supersymmetry is broken today.  On the contrary, it
 causes a severe disaster since breaking supersymmetry 
inevitably induces  a
positive vacuum energy density with SUSY breaking scale, 
$\rhov \sim (1{\rm TeV})^4 \sim 10^{29}{\rm gcm^{-3}}$ or larger.

This disaster is cured by making SUSY local, namely, introducing
supergravity (SUGRA).  In SUGRA we do not have a relation like (\ref{susy})
and the vacuum energy density can be either negative or positive. 
Usually it is negative due to a ``$-3|W|^2$'' term in the scalar potential
where $W$ is the superpotential.  Then we may incorporate a positive
contribution from SUSY breaking to fine-tune the resultant vacuum energy
density vanishing.  To this end, SUGRA does not solve the conceptual
Problem I either.

Thus SUSY/SUGRA does not help after all and no other symmetry is known
to realize vanishing vacuum energy density even after SUSY breaking.

\subsection{Adjustment mechanism}

This is an attempt to dynamically cancel the cosmological constant
toward zero as the passage of time.  In 1982 Dolgov proposed a model
with a massless scalar field nonminimally coupled to the scalar
curvature \cite{Dolgov}.  
He showed that in this model the effective cosmological
constant decreases in the manner inversely proportional to the square of
cosmic time toward zero no matter how large it may be at the outset.
This evolution law is the same as that in scalar-tensor model of Fujii
\cite{Fujii} proposed in the same year. 
The problem with Dolgov's model is the effective gravitational constant
also decreases toward zero at the same time, so it cannot serve as the
solution in the real world.

As for the impossibility of finding a  successful adjustment mechanism
to the Problem I, there is a no-go theorem of Weinberg \cite{Weinberg} 
which states
that one cannot find an 
equilibrium solution of the field
equations with vanishing cosmological constant without fine-tuning.
As a simple example, consider a model with $N$ scalar degrees of 
freedom $\phi_i$.  Then an equilibrium or stationary configuration
is determined from $N$ equations such as $\partial_{\phi_i}V[\phi]=0$
for each $i=1,...,N$ where $V$ represents the potential of the system under
consideration.  In order to realize vanishing vacuum energy at the
equilibrium configuration, another equation should be satisfied there 
to ensure $\Lambda=0$, which requires a fine-tuning because we have
already used $N$ degrees of freedom to specify the equilibrium
configuration.   

\subsection{Quantum Cosmological Approach}
This approach was quite fashionable more than a decade ago.
It is based on the wave function of the universe.  In Hartle-Hawking's
approach \cite{Hartle:1983ai}, 
the wave function with three geometry $h_{ij}$ and matter
configuration $\phi$ is expressed by an Euclidean path integral over all
the compact manifolds $C'$ with boundary $(h_{ij},\phi)$,
\beq
\Psi _{HH}\left[ {h_{ij},\phi } \right]=\int_{C'} {[dg][d\phi
]e^{-I[g,\phi ]}}.
\eeq
In this prescription, the
 expectation value of an operator $Y$ is given by
a path integral over all the compact Euclidean manifolds $C$
as
\beq
\left\langle Y \right\rangle _{HH}={{\left( {\Psi _{HH},Y\ \Psi _{HH}}
\right)} \over {\left( {\Psi _{HH},\Psi _{HH}} \right)}}={{\int_C
{[dg][d\phi ]Ye^{-I[g,\phi ]}}} \over {\int_C {[dg][d\phi ]e^{-I[g,\phi
]}}}}.
\eeq
If this path integral is dominated by the de Sitter instanton with the
action 
\beq
I=-{1 \over {16\pi G}}\int {\left( {R-2\Lambda } \right)}\sqrt
gd^4x=-{{3\pi } \over {G\Lambda }},
\eeq
the wave function is peaked at $\Lambda=0$.
\beq
\Psi _{HH}\left[ {h_{ij},\phi } \right]=\int_{C'} {[dg][d\phi ]e^{-I[g,\phi ]}}\propto \exp \left( {{{3\pi } \over {G\Lambda }}} \right),
\eeq
suggesting the vanishing cosmological constant \cite{Baum:mc}.

Later on Coleman proposed to incorporate sum over topology with wormhole
configurations connected to mother and baby universes.  Then all the
physical constants become a variable dependent on the state of baby
universes $\alpha$ and the expectation value reads,
\beq
\left\langle Y \right\rangle ={{\int {[d\alpha ]e^{-{\textstyle{{\alpha
^2} \over 2}}}\left\langle Y \right\rangle _\alpha Z\left( {\left\{
\alpha  \right\}} \right)}} \over {\int {[d\alpha
]e^{-{\textstyle{{\alpha ^2} \over 2}}}Z\left( {\left\{ \alpha
\right\}} \right)}}},\ \ \ Z\left( {\left\{ \alpha  \right\}}
\right)\propto \exp \left[ {\exp \left( {{{3\pi } \over {G\Lambda (\{
\alpha \} )}}} \right)} \right]. \label{col}
\eeq
Thus we find an even sharper peak at $\Lambda=0$ \cite{Coleman:1988tj}.
This mechanism may also be used to fix other physical parameters.  
For example, strong CP problem may be solved without introducing
axions \cite{Nielsen:1988kf}.  

In these Euclidean approaches there is no time dependence in the
expressions of expectation values and they should rather be interpreted
as an average expectation value throughout the possible cosmic history
which would coincide with that in the most stable ground state \cite{Weinberg}.

There are a number of serious problems in these approaches.  First,
the Euclidean path integral is not well-defined,
because, once gravity is included, the action is 
not positive definite even after Wick rotation.  Second, no conserved
and positive-definite probability current is known in this approach,
which makes its interpretation difficult.  
Third, in the original calculation of wormhole and baby universe system,
phase of sum over spheres was miscounted and the sharper peak observed
in (\ref{col}) diminishes in the correct calculation \cite{Polchinski:ua}.

For these reasons, although this approach is attractive in that it is a
quantum theory, this solution is not taken seriously now. 

\subsection{Higher dimensional models}

Higher dimensional theories may open up a new perspective to the
cosmological constant problem, because lower dimensional subspace may
have flat geometry even if energy-momentum tensor is nonvanishing.
Hence if we could succeed in removing the gravitational effect of vacuum
energy on cosmic expansion of our three-brane, this could serve as a
solution to the problem.  For example, if we found a modified Friedmann
equation of our three-brane like
\beq
H^2  = f\left( {\rho ,p} \right)\left( {\rho  + p} \right) + ...,
\eeq
after an appropriate dimensional reduction, such an equation would  
allow for the Minkowski solution $H=0$ in the presence of arbitrary
vacuum energy density with $\rho _{\rm v}=-p_{\rm v}$ on the three-brane.
While many ideas have been proposed these days I do not go in detail of
them because there will be many brane world talks in this workshop and I
hope this issue will be covered there.  It seems, however, many of them
require fine-tuning, some of which are as severe as the original one, to
realize such an effective equation on the brane so they could hardly
evade the no-go theorem. 

I would like to briefly mention, however, an exception which makes use
of infinite-volume extradimensions with large-distance modification of
gravity on the brane proposed by Dvali, Gabadadze, and 
Shifman \cite{Dvali:2002fz}. They
start with an $4+N$ dimensional action,
\beq
S = M_*^{2 + N} \int {d^4 } xd^N y\sqrt G R_{4 + N}  
+ \int {d^4 }x
\sqrt g \left( {M_G^2 R_4  + {\cal L}_{{\rm{SM}}}  +\Lambda_4 } \right),
\eeq
where $M_*$ is the fundamental Planck scale and $M_G$ is the four
dimensional counterpart induced by quantum correction of matter in four
dimension with the Lagrangian $\cal{L}_{{\rm{SM}}}$.
In this model the four dimensional graviton $h_{\mu\nu}$ satisfies a
modified equation,
\beq
\left[ {1 + \frac{{F_N }}{{r_c^2\square }}} \right]\square
h_{\mu \nu }  = 0,
\eeq
where $F_N$ is a constant and
$r_c  \equiv {{M_G }}/{{M_*^2 }}$ is the crossover distance beyond
which gravity becomes weak. It should satisfy $r_0 \gtrsim
H_0^{-1}\approx 10^{28}$cm to ensure standard cosmology.  Hence we
should take $M_* \lesssim 10^{-3}$eV.  Since the cosmological constant
problem is a far-infrared (large-scale) problem, four dimensional
curvature is insensitive to it in such a modified theory of gravity.
Thus we may naturally have an almost flat four dimensional spacetime no
matter how large $\Lambda_4$ may be.  Although 
this model relies on a huge
hierarchy  $M_*/M_G \lesssim 10^{-30}$, it is stable under the
assumption of unbroken supersymmetry in the bulk.

\subsection{Anthropic principle}
Since the cosmological constant problem is such a difficult problem to
solve, it is natural that many people are tempted to resort its solution
to the anthropic consideration in which one attempts to interpret 
natural phenomena based on our existence \cite{Barrow}.

Consider a proposition: We, human being, cannot live in a universe with
a large cosmological constant, positive or negative.  We can easily
convince ourselves that this proposition is correct at least
qualitatively, and more quantitative proofs can be found in the
literature \cite{ant}.  
Now consider its contraposition: The cosmological constant
in our Universe is small, because human being exists.
Since a proposition and its contrapositive always have the same truth
value, once we convince ourselves that the above proposition is correct,
it can serve as a solution to the cosmological constant problem at least
logically.

But this fact alone does not guarantee that the anthropic consideration
is useful in theoretical physics.  Let us consider a perhaps more
illustrative question: Why does our Universe have three spatial
dimensions?   A possible answer based on the anthropic principle is that
if the spatial dimension is not equal to three, an orbit of a planet is
not closed, so that its climate would be too unstable to accommodate
human life.  Again this solution is logically correct answer to the
original question, but it is clear that such an answer provides no
insights for the compactification mechanism of extra dimensions which is
one of the central issue in higher-dimensional theories.  Thus in this
sense this approach does not contribute to progress in theoretical
physics.  

I do not claim that we should not use the anthropic principle at all.
On the contrary, I admit that this could be useful at some stage of
investigation.  As for the cosmological constant, we might use it once
we have reached an ultimate theory which clarifies what physics controls
it but does not predict of its value itself.  

The anthropic principle
resembles {\it sake}; if you use an appropriate amount at night, you can
be happy, but if you drink it too much in the morning, you will lose
your job.  I think it is clear that the situation of investigation of
cosmological constant problem is yet in the morning or in the premature
stage and use of the anthropic principle now could be dangerous because
we might well miss a chance to reach an essence of theoretical physics.

Nevertheless there is an exception.  We may or even should resort the
solution of what is called the coincidence problem or why-now problem,
namely the question why vacuum energy is dominating only recently, 
to the anthropic principle, because ``now'' is defined by our very
existence.  

\section{Proposed solutions to the Problem II}

I now move on to  various proposals to explain the origin of the tiny
dark energy density observed.  They are based on a common assumption
that there exists some ground state with a vanishing vacuum energy
density where the Problem I is solved.  I would also like to note that
two of the suggested solutions to the Problem I, namely the anthropic
consideration \cite{ant} and Dvali et al's higher dimensional model
\cite{Dvali:2002fz},
 may also naturally
accommodate a tiny nonvanishing dark energy density.  

\subsection{Decaying $\Lambda$ or quintessence}

This is an attempt to identify the vacuum-like energy density with a
dynamical, slowly-changing and possibly spatially inhomogeneous
component with negative pressure \cite{Dolgov,Fujii,Ratra,Sato}.  
Typical example is a scalar field $Q$, now called quintessence, 
slowly rolling down its extremely flat potential \cite{Ratra,quint}.  
By virtue of its time
dependence, the Problem II, or why the observed value of the dark energy today
is unnaturally small compared with, say, the Planck scale, may be
relaxed.  In fact, by choosing an appropriate shape of the potential
$V[Q]$ it has been advocated that one can find a natural solution to the
coincidence problem in which the energy density of the
quintessence field decreases during radiation domination tracking
evolution of the total energy density and becomes dominant some time
after the equality time.  Such a tracker solution is possible with a
potential like $V[Q]=M^4(M_G/Q)^n$ with $n>0$ or 
$
V\left[ Q \right] = M^4 \left[ {\exp \left( {{{M_G } \mathord{\left/
 {\vphantom {{M_G } Q}} \right.
 \kern-\nulldelimiterspace} Q}} \right) - 1} \right],
$
both with a canonical kinetic term.  By nature of the tracker solution,
the scalar field is still evolving today in the time scale of cosmic
expansion, which means that the equation of state satisfies
\[
w_Q  = \frac{{p_Q }}{{\rho _Q }} = \frac{{{{\dot Q^2 } \mathord{\left/
 {\vphantom {{\dot Q^2 } {2 - V[Q]}}} \right.
 \kern-\nulldelimiterspace} {2 - V[Q]}}}}{{{{\dot Q^2 } \mathord{\left/
 {\vphantom {{\dot Q^2 } {2 + V[Q]}}} \right.
 \kern-\nulldelimiterspace} {2 + V[Q]}}}} \gtrsim  - 0.75,
\]
and that its mass is comparable to the Hubble parameter today,
$m_Q=\sqrt{V''[Q]}\approx H_0=10^{-33}$eV.

As will be seen below the above characteristic of the equation of state
is now confronting with observational tests.  Furthermore from
theoretical view point it is difficult to build a sensible particle
physics model to realize such a small effective mass with a huge
hierarchy $m_Q/M_w=10^{-44}=10^{-26}M_w/M_G$, where $M_w$ is the weak
scale.  Nevertheless such a small mass might
be stably realized if protected by some symmetry.  One example is the
model proposed by Nomura, Watari, and Yanagida \cite{Nomura:2000yk}
in which the quintessence
field is identified with a pseudo Nambu-Goldstone boson.  This model,
unfortunately, does not lead to a tracker behavior.

\subsection{Kinetically driven quintessence}

It was observed by Armend\'ariz-Pic\'on, Damour, and Mukhanov that 
a scalar field can drive inflation even if it has no potential provided
that it has an appropriate non-canonical kinetic term
\cite{Armendariz-Picon:1999rj}.  This idea was first applied to explain
the accelerated expansion observed today by Chiba, Okabe, and Yamaguchi
under the name of kinetically driven quintessence \cite{Chiba:1999ka}.  
Now it is abbreviated as k-essence
\cite{Armendariz-Picon:2000dh,Armendariz-Picon:2000ah}.
The evolution of the equation of state, $w$, is highly nontrivial in
this class of models and may be observationally tested.
An example of the scalar Lagrangian giving rise to a desired recent cosmic
acceleration is given by \cite{Armendariz-Picon:2000dh}
\beq
L = \frac{1}{{\varphi ^2 }}\left( {2\sqrt {1 + X}  - 2.01 + 0.03\left( {10^{ - 5} X} \right)^3  - \left( {10^{ - 6} X} \right)^4 } \right),{\rm{  }}X \equiv  - \frac{1}{2}\left( {\partial \varphi } \right)^2  = \frac{1}{2}\dot \varphi ^2 .
\eeq

\subsection{QCD trace anomaly}
 
The QCD trace anomaly contributes to the vacuum energy density
with the amount
\beq
\left\langle {F_{\mu \nu }^a F_a^{\mu \nu } } \right\rangle  =
\order\left( {\Lambda _{\rm QCD}^4 } \right) \approx 
\order\left( ({100{\rm{MeV}}})^4 \right). 
\eeq
Suppose that this contribution is canceled by the intrinsic
cosmological constant in Minkowski vacuum state which is a part of the
usual assumption that the Problem I is solved in the ground state.
Then Sch\"utzhold \cite{Schutzhold:pr} 
calculates the same quantity in the Friedmann vacuum
under this assumption and finds a weakly time-dependent dark energy density
\beq
\rho _{{\rm{v}}}  = \order\left( {\Lambda _{\rm QCD}^3 H} \right).
\eeq
Although this proposal is attractive in that it is a result of serious
calculation of quantum field theory in the curved spacetime, the
resultant value might be too large, 
${\Lambda _{\rm QCD}^3 H_0}\gtrsim 30\rhocrz$ for 
$\Lambda _{\rm QCD}\gtrsim 100$MeV. 

\subsection{False vacuum energy}

All the proposals to the problem II discussed so far predict
time-dependent dark energy.  But we should also consider the possibility
that we live in a false vacuum state with a finite vacuum energy
density.  Most people believe that such a possibility is unnatural
due to the smallness of the observed vacuum energy, 
$\rhov=10^{-120}M_G^4=({\rm meV})^4$.  Several attempts do exist,
however, to explain the existence of a false vacuum state with the
desired magnitude of vacuum energy density 
naturally by introducing some discrete symmetry
\cite{Garretson:1993kg} or global symmetry \cite{Barr:2001vh}
and then breaking it with some appropriate nonrenormalizable terms.

\subsection{Non-perturbative explanation with degenerate vacua}

Another possibility that predicts  time-independent dark energy
without introducing any tiny numbers is a theory with degenerate
perturbative vacua.  It starts with the observation
that the cosmological constant Problem I is a perturbative problem, that
is, why we observe vanishingly small vacuum energy even after
perturbative quantum correction, (\ref{qft}), and we assume that this
problem is solved by a yet unknown mechanism as assumed in all the
other proposals above.  If there exist two or more degenerate
perturbative vacua, say, $|+\rangle$ and $|-\rangle$ whose vacuum energy
is vanishing under the assumption that the perturbative
Problem I is solved there, and if quantum tunneling between perturbative
vacua is possible, these states are no longer the real ground state. 
Including the effect of quantum tunneling, which we assume is described
by an instanton with the Euclidean action $S_0$, the energy eigen
states are given by superposition of perturbative vacua, 
\beq
\left| S \right\rangle ={1 \over {\sqrt 2}}\left( {\left| + \right\rangle \ +\ \left| - \right\rangle } \right),~~~\left| A \right\rangle ={1 \over {\sqrt 2}}\left( {\left| - \right\rangle \ +\ \left| - \right\rangle } \right),
\eeq
where $|S\rangle$ is the real ground state with $\rhov=-m^4e^{-S_0}$
and the first excited state is $|A\rangle$ with $\rhov=m^4e^{-S_0}$,
with $m$ being the energy scale associated with the quantum transition.
Thus each perturbative vacua is expressed by the superposition of energy
eigenstates with exponentially small energy gap, and if we still live in
one of the perturbative vacua we observe a finite magnitude of dark
energy density $\rhov=m^4e^{-S_0}$ with the probability of 50\% in this
case.  In order to match the predicted value with the observed value, we
should take $S_0=120\ln10+4\ln(m/M_G)$, and the longevity of the
perturbative vacua requires $m \gtrsim M_G$
\cite{Yokoyama:2001ez}.  

This solution is attractive in two aspects:
one is that the Problems I and II are properly separated and the other
is that, as mentioned above, it involves no small numbers and the
smallness of the observed dark energy is the largeness of the instanton
action.  The former feature may particularly be important because in
other proposals no explanations have been given why {\it only} the
contribution of
quintessence or k-essence fields to the vacuum energy density
remains finite after solving the Problem I.

\section{Observational constraints on dark energy}
In order to examine which, if any, of the above proposals is the right
one, it is important to observationally constrain the equation of state,
$w=p/\rho$, of dark energy.  An increasing number of observational
analysis have been done these days.

For example, Corasaniti and Copeland used SNIa data and the location of
the three peaks of CMB anisotropy and obtained an ambitious constraint
$-1\leq w \leq -0.93$ \cite{Corasaniti:2001mf}.

Melchiorri, on the other hand, performed a joint analysis of CMB, SNIa,
HST, and the large-scale structures and concluded $w \leq -0.85$ at
$1\sigma$ and $w \leq -0.72$ at $2\sigma$  \cite{Melchiorri:2002yy}.
His result is perfectly consistent with the cosmological constant
$w=-1$.

Percival et al.\ finds from 2dF data combined with CMB anisotropy $w
\leq -0.51$ at $2\sigma$ and the probability distribution of $w$ is
peaked around $w= -1$, namely, they favor the cosmological constant
\cite{Percival}. 

From REFLEX sample of the redshift distribution of X-ray clusters
combined with SNIa data, Schuecker et al.\ find 
$w=-0.95^{+0.30}_{-0.35}$ under the assumption of spatially flat Universe
\cite{Schuecker:2002yj} and claims that the
most natural interpretation of the data is the cosmological constant.

Thus we may conclude that the tracker solution of the quintessence
scenario is
already in a difficult situation and that currently available data seem
to favor the cosmological constant.
Note that CMB data used above are pre-WMAP ones.  The new constraint
obtained with WMAP is 
$w<-0.78$ at $2\sigma$ \cite{wmap}.

\section{The future of the Universe dominated by $\Lambda$}

If our Universe is dominated by time-independent vacuum energy density
as suggested by numerous observations and if it is stable against phase
transitions, we can draw some interesting conclusions about the future
of our Universe \cite{Starobinsky:1999yw,Loeb:2001dh}. 

The existence of a positive cosmological constant means that our
Universe is in a stage of asymptotically de Sitter expansion.
So there will be an event horizon at the distance $\approx 5.1$ Gpc
and we cannot go to any galaxies with $z > 1.8$ not only in practice but
also in principle.  We can observe a galaxy at $z=5$ only for 6.4 Gyr
from now.

Future evolution of nearby large-scale structure in our Universe dominated
by a cosmological constant has been studied by Nagamine and Loeb using
an $N$-body simulation \cite{Nagamine:2002wi}.
They find:
\begin{itemize}
\item Evolution of large-scale structure continues for 28 Gyrs.
\item Our local group will not be bound to the Virgo cluster.
\item Our galaxy is likely to merge with Andromeda galaxy within the
      Hubble time.
\item This will be the only galaxy within the horizon 100 Gyr later.
\end{itemize}

This means that extragalactic astronomy and cosmology must be solved
within 100 Gyr.  So we should hurry up!


\begin{thebibliography}{99}
\bibitem{Ein}A.\ Einstein, Preuss.\ Akad.\ Wiss.\ Berlin Sitzer. (1917) 142.
\bibitem{Hubble}E.P.\ Hubble, Proc.\ Nat.\ Acad.\ Sci. {\bf 15} (1929) 168.
\bibitem{Weinberg}S.\ Weinberg, Rev.\ Mod.\ Phys. {\bf 61} (1989) 1.
\bibitem{SN}S.\ Perlmutter et al.,  ApJ, {\bf 517} (1999) 565;
A.G.\ Riess  et al.,  AJ, {\bf 116} (1998) 1009.
\bibitem{hst}W.~L.~Freedman {\rm et al.},
ApJ.\  {\bf 553} (2001) 47.
\bibitem{SZ} E.~D.~Reese, J.~E.~Carlstrom, M.~Joy, J.~J.~Mohr, L.~Grego and W.~L.~Holzapfel,
ApJ\  {\bf 581} (2002) 53.
\bibitem{Olive}R.H.\ Cyburt, B.D.\ Fields, and K.A.\ Olive,
	arxiv:astro-ph/0302431. 
\bibitem{Carlberg}
R.G.\ Carlberg, H.K.C.\ Yee, and E.\ Ellingson, ApJ {\bf 478} (1997) 462. 
\bibitem{White}D.A.\ White, A.C.\ Fabian, MNRAS {\bf 273} (1995) 73.
\bibitem{Dodds}J.A.\ Peacock and S.J.\ Dodds, MNRAS {\bf 267} (1994) 1020
\bibitem{cc} P.\ de Bernardis et al., Nature {\bf 404}(2000) 955;
S.\ Hanany et al., ApJ {\bf 545} (2000) L5.
\bibitem{Jaffe:2000tx}
A.~H.~Jaffe {\rm et al.}  [Boomerang Collaboration],
Phys.\ Rev.\ Lett.\  {\bf 86} (2001) 3475.
\bibitem{S}A.\ Sandage and G.A.\ Tammann, ApJ {\bf 256} (1982) 339. 
\bibitem{dV}G.\ de Vaucouleurs and H.G.\ Corwin jr., ApJ {\bf 297}
	(1985) 23.
\bibitem{wmap}C.L.\ Bennett et al., arXiv:astro-ph/0302207; D.N.\ Spergel et
	al., arXiv:astro-ph/0302209.
\bibitem{Dolgov}A.D.\ Dolgov, In ``Very Early Universe'' Eds.\ G.W.\
	Gibbons, S.W.\ Hawking, and S.T.\ Siklos, (Cambridge 1982) 449.
\bibitem{Fujii}Y.\ Fujii, Phys.\ Rev.\ {\bf D26} (1982) 2580.
\bibitem{Hartle:1983ai}
J.B.~Hartle and S.W.~Hawking,
Phys.\ Rev.\ D {\bf 28} (1983) 2960.
\bibitem{Baum:mc}
E.~Baum,
Phys.\ Lett.\ B {\bf 133} (1983) 185; 
S.W.~Hawking,
Phys.\ Lett.\ B {\bf 134} (1984) 403.
\bibitem{Coleman:1988tj}
S.~R.~Coleman,
Nucl.\ Phys.\ B {\bf 310} (1988) 643.
\bibitem{Nielsen:1988kf}
H.~B.~Nielsen and M.~Ninomiya,
Phys.\ Rev.\ Lett.\  {\bf 62} (1989) 1429.
\bibitem{Polchinski:ua}
J.~Polchinski,
Phys.\ Lett.\ B {\bf 219} (1989) 251.
\bibitem{Dvali:2002fz}
G.~Dvali, G.~Gabadadze and M.~Shifman,
arXiv:hep-th/0208096.
\bibitem{Barrow}For a review, see {\it e.g.} J.D.\ Barrow and F.J.\
	Tipler, ``The Anthropic Cosmological Principle'' (Oxford 1986).
\bibitem{ant}S.\ Weinberg, Phys.\ Rev.\ Lett. {\bf 59} (1987) 2607;
G.\ Efstathiou, MNRAS {\bf 274} (1995) L73;J.~Garriga and A.~Vilenkin,
Phys.\ Rev.\ D {\bf 61} (2000) 083502.
\bibitem{Ratra}
B.~Ratra and P.~J.~Peebles,
Phys.\ Rev.\ D {\bf 37} (1988) 3406.
\bibitem{Sato}K.\ Sato, N.\ Terasawa, and J.\ Yokoyama, in ``The Quest
	for the Fundamental Constants in Cosmology'' Eds.\ J.\ Audouze
	and J.\ Tran Thanh Van (Editors Frontieres, 1989) 193.
\bibitem{quint} I.\ Zlatev, L.\ Wang, and P.J.\ Steinhardt, Phys.\ Rev.\
	Lett. {\bf 82} (1999) 896; P.J.\ Steinhardt, L.\ Wang, and I.\
	Zlatev, Phys.\ Rev.\ {\bf D59} (1999) 23504.
\bibitem{Nomura:2000yk}
Y.~Nomura, T.~Watari and T.~Yanagida,
Phys.\ Lett.\ B {\bf 484} (2000) 103.
\bibitem{Armendariz-Picon:1999rj}
C.~Armendariz-Picon, T.~Damour and V.~Mukhanov,
Phys.\ Lett.\ B {\bf 458} (1999) 209.
\bibitem{Chiba:1999ka}
T.~Chiba, T.~Okabe and M.~Yamaguchi,
Phys.\ Rev.\ D {\bf 62} (2000) 023511.
\bibitem{Armendariz-Picon:2000dh}
C.~Armendariz-Picon, V.~Mukhanov and P.~J.~Steinhardt,
Phys.\ Rev.\ Lett.\  {\bf 85} (2000) 4438.
\bibitem{Armendariz-Picon:2000ah}
C.~Armendariz-Picon, V.~Mukhanov and P.~J.~Steinhardt,
Phys.\ Rev.\ D {\bf 63} (2001) 103510.
\bibitem{Schutzhold:pr}
R.~Schutzhold,
Phys.\ Rev.\ Lett.\  {\bf 89} (2002) 081302.
\bibitem{Garretson:1993kg}
W.~D.~Garretson and E.~D.~Carlson,
Phys.\ Lett.\ B {\bf 315} (1993) 232.
\bibitem{Barr:2001vh}
S.~M.~Barr and D.~Seckel,
Phys.\ Rev.\ D {\bf 64} (2001) 123513.
\bibitem{Yokoyama:2001ez}
J.~Yokoyama,
Phys.\ Rev.\ Lett.\  {\bf 88} (2002) 151302;
Int.\ J.\ Mod.\ Phys.\ D {\bf 11} (2002) 1603.
\bibitem{Corasaniti:2001mf}
P.~S.~Corasaniti and E.~J.~Copeland,
Phys.\ Rev.\ D {\bf 65} (2002) 043004.
\bibitem{Melchiorri:2002yy}
A.~Melchiorri,
arXiv:astro-ph/0204262.
\bibitem{Percival}
W.J.\ Percival et al.\ MNRAS, {\bf 337} (2002) 1068.
\bibitem{Schuecker:2002yj}
P.~Schuecker, R.~R.~Caldwell, H.~Bohringer, C.~A.~Collins and L.~Guzzo,
arXiv:astro-ph/0211480.
\bibitem{Starobinsky:1999yw}
A.~A.~Starobinsky,
Grav.\ Cosmol.\  {\bf 6} (2000) 157.
\bibitem{Loeb:2001dh}
A.~Loeb,
Phys.\ Rev.\ D {\bf 65} (2002) 047301.
\bibitem{Nagamine:2002wi}
K.~Nagamine and A.~Loeb,
arXiv:astro-ph/0204249.

\end{thebibliography}
\end{document}